\documentclass[aps,prl,reprint,superscriptaddress,nofootinbib,floatfix]{revtex4-2}
\usepackage{amsmath,amssymb,bm}
\usepackage{graphicx}
\usepackage{xcolor}
\usepackage[colorlinks=true,citecolor=blue!55!black,urlcolor=blue!55!black,linkcolor=blue!55!black]{hyperref}
\graphicspath{{figures/}}
\hypersetup{pdftitle={Universal Leading Quantum Stress Near the Schwarzschild Singularity},pdfauthor={Shun Jiang and Jie Jiang}}
\begin{document}
\raggedbottom
\title{Universal Leading Quantum Stress Near the Schwarzschild Singularity}
\author{Shun Jiang}
\affiliation{School of Physics and Optoelectronics, South China University of Technology, Guangzhou 510641, China}
\author{Jie Jiang}
\email{jiejiang@bnu.edu.cn}
\affiliation{Faculty of Arts and Sciences, Beijing Normal University, Zhuhai 519087, China}

\begin{abstract}
Determining the local quantum source near a spacelike singularity is essential for assessing semiclassical backreaction inside black holes. By standard point splitting, we provide, to our knowledge, the first complete numerical determination of the renormalized stress-energy tensor (RSET) of a massless scalar field at arbitrary curvature coupling in the four-dimensional Schwarzschild interior. Calculations in the Unruh and Hartle--Hawking states reach $r/M\simeq10^{-4}$, yielding a common leading $r^{-6}$ coefficient tensor and coupling-dependent anisotropic pressure ratios. Uniform estimates of the complete mode sums then establish a rigidity theorem: at each fixed coupling, any Hadamard-state change on the common interior evolution domain contributes only subleading stress. The leading tensor is therefore insensitive to time and directional dependence of the quantum environment, in contrast with state-dependent leading Cauchy-horizon coefficients in asymptotically flat Reissner--Nordstr\"om spacetime. Its directional null projections define positive, mixed-sign, and negative regimes in coupling and direction, specifying the local mean focusing source; at conformal coupling, all leading projections are positive.
\end{abstract}
\maketitle

\textit{Introduction.---}
Quantum fields can profoundly affect the interior structure of black holes. Their renormalized stress-energy tensor (RSET) supplies the mean source in the semiclassical Einstein equation and is central to assessing the fate of classical interior geometries. Studies of inner-horizon quantum fluxes~\cite{ZilbermanLeviOri2020,ZilbermanKerr2022}, charge transport~\cite{Klein:2021ctt}, and infinite quantum twisting~\cite{Klein:2024sdd} have revealed singular stresses and currents with potentially decisive consequences for the inner geometry. These advances motivate a corresponding description of the Schwarzschild interior, which terminates at a spacelike curvature singularity. Determining the complete quantum stress in this high-curvature region is essential for assessing semiclassical backreaction on its singularity structure.

Covariant point splitting~\cite{Christensen1976,Christensen1978} and its numerical implementations, including pragmatic mode-sum regularization and angular splitting~\cite{LeviOri2015,Levi:2016esr,Levi2017}, provide a systematic route to the RSET. Numerical work has determined stresses in the Schwarzschild exterior~\cite{Candelas1980,HowardCandelas1984,TaylorBreenOttewill2022} and vacuum polarization in the interior~\cite{CandelasJensen1986,LanirLeviOri2018}. In particular, the interior calculation of Lanir, Levi, and Ori covered $M\le r<2M$~\cite{LanirLeviOri2018}; it did not determine the complete RSET deep near the singularity. Analytic treatments anticipated an $r^{-6}$ divergence and estimated its coefficients~\cite{Anderson1995,Hiscock:1997jt,Hamilton:2016iid}. The exponent alone, however, does not specify the physical source. The leading coefficient matrix fixes its amplitude, energy density, and anisotropic pressures, thereby determining the strength and directional character of the stress entering semiclassical gravity. A controlled numerical determination of this matrix is therefore crucial to understanding the quantum source near the singularity.

Astrophysical black holes are generally exposed to surrounding radiation rather than an idealized vacuum environment. Cosmic expansion changes the temperature of background radiation~\cite{Li:2021hxe}, motion relative to the radiation introduces directional dependence~\cite{Ferreira:2020aqa}, and passage through spatially inhomogeneous radiation regions changes the incident spectrum and its time dependence. The resulting incoming-state boundary conditions need not be stationary or spherically symmetric and may contain nontrivial correlations between modes. Whether this environmental dependence survives in the leading quantum stress inside a black hole is therefore an important question. In asymptotically flat Reissner--Nordstr\"om spacetime, the Unruh and Hartle--Hawking states have different coefficients of the leading Cauchy-horizon stress divergence~\cite{ZilbermanLeviOri2020}, showing that environmental dependence can enter the dominant term. Near the Schwarzschild spacelike singularity, however, quantum-field evolution accompanies unbounded growth of the background curvature, a mechanism distinct from the infinite blueshift at a Cauchy horizon. Does the leading quantum stress near this spacelike singularity likewise retain information about the surrounding quantum environment?

In this Letter, we numerically determine all independent RSET components at arbitrary curvature coupling in the Schwarzschild interior for the Unruh and Hartle--Hawking states. To our knowledge, this is the first complete interior RSET calculation by standard point splitting; it reaches $r/M=10^{-4}$ and resolves the same leading coefficient matrix in both states. The full coupling dependence follows within one calculation of the stress operator and vacuum-polarization derivatives. We then combine the smoothness of Hadamard-state differences with uniform energy estimates of the full mode expansion to prove that state changes on a common interior evolution domain produce only subleading stress, without equilibrium, state-symmetry, or finite-particle assumptions. Thus, at fixed coupling, the numerically determined leading tensor remains rigid even in nonstationary or nonspherically symmetric external quantum environments.

\begin{figure*}[t]
\centering
\includegraphics[width=0.98\textwidth]{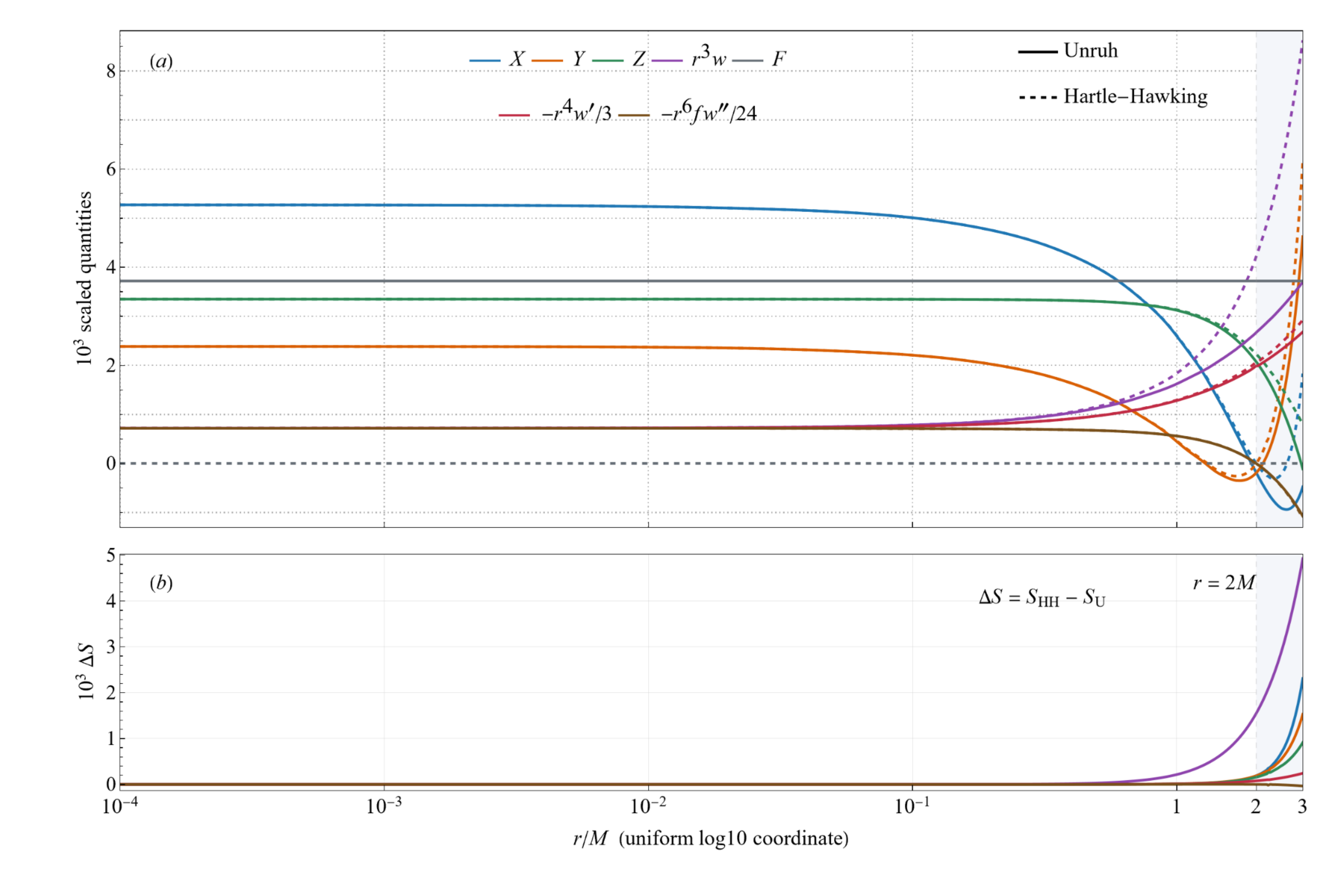}
\caption{Directly computed basis for the complete RSET at arbitrary curvature coupling, with $M=1$. (a) Coupling-independent stress-basis quantities $X=-r^7fT^{(0)t}{}_t$, $Y=r^7fT^{(0)r}{}_r$, $Z=r^6T^{(0)\theta}{}_\theta$, and $F=-200\pi r^2T^{(0)r}{}_t$, together with the scalar profiles $r^3w$, $-r^4w'/3$, and $-r^6fw''/24$. All ordinates include the factor $10^3$. Solid curves denote Unruh and dashed curves Hartle--Hawking; shading marks the exterior. The corresponding diagonal-stress and scalar curves in the two states approach the same respective plateaus. The three scalar profiles share the asymptotic value $q$, and the stress plateaus determine the complete leading tensor. (b) Hartle--Hawking minus Unruh differences of the six nonflux quantities, with the colors of (a). The vanishing small-$r$ limits show the common leading coefficients of the two states. Interior curves use angular splitting; exterior curves use the validated $t$-splitting results described in the text. Curves interpolate the numerical profiles.}
\label{fig:basis}
\end{figure*}

\textit{Numerical RSET at arbitrary curvature coupling.---}
We consider a massless real scalar field with curvature coupling $\xi$ on the Schwarzschild geometry,
\begin{equation}
 ds^2=-f\,dt^2+f^{-1}dr^2+r^2d\Omega^2,\qquad
 f=1-\frac{2M}{r}.
 \label{eq:metric}
\end{equation}
We set $G=c=\hbar=M=1$ below and write $T^{(\xi)}_{ab}=\langle T^{(\xi)}_{ab}\rangle_{\rm ren}$. Ricci flatness makes the wave equation $(\Box-\xi R)\Phi=0$ independent of $\xi$, so the same scalar modes enter the calculation at every coupling; the coupling dependence resides in the stress operator. For the same state and matched renormalization prescription,
\begin{equation}
 T^{(\xi)}_{ab}=T^{(0)}_{ab}
       +\xi(g_{ab}\Box-\nabla_a\nabla_b)w.
 \label{eq:improvement}
\end{equation}
This expression follows from the complete conserved point-split operators and Synge's rule~\cite{Moretti:2001qh,Hack:2012qf,Zahn:2025tnh}; its derivation is given in Sec.~II of the Supplemental Material (SM)~\cite{Supplement}. For the stationary, spherically symmetric states computed here, Eq.~\eqref{eq:improvement} organizes the full RSET calculation into a common numerical basis: the $\xi$-independent stress contribution $T^{(0)}_{ab}$, the vacuum polarization $w=\langle\Phi^2\rangle_{\rm ren}$, and its radial derivatives $w',w''$. These quantities are all evaluated directly by numerical point splitting and together determine every independent RSET component at arbitrary fixed $\xi$.

Inside the horizon we use the angular-splitting scheme of pragmatic mode-sum regularization~\cite{LeviOri2015,Levi:2016esr,Levi2017}, together with the interior state representation of Refs.~\cite{LanirLeviOri2018,Lanir:2017oia}, to determine the stress and vacuum-polarization basis entirely through numerical mode sums. Outside the horizon we use standard $t$-splitting results, cross-checked against numerical data extracted from the published vector figures of Ref.~\cite{Levi2017}. Renormalization uses the Christensen subtraction terms appropriate to each splitting direction~\cite{Christensen1976,Christensen1978}; radial derivatives of the vacuum polarization include the corresponding derivatives of the subtraction terms. Computational details, mode convergence, exterior-data checks, and independent conservation and trace-identity checks are given in SM Secs.~II--V~\cite{Supplement}.

Figure~\ref{fig:basis} displays the numerical basis from $r=3$ to $10^{-4}$. Near the singularity, the Unruh and Hartle--Hawking curves approach the same plateau for each rescaled diagonal-stress or scalar quantity. The stress plateaus are consistent with leading $r^{-6}$ behavior and satisfy $X\to2\lambda_t$, $Y\to-2\lambda_r$, and $Z\to\lambda_\perp$. The three independently computed scalar curves $r^3w$, $-r^4w'/3$, and $-r^6fw''/24$ also share a common plateau $q$. Their normalization factors account for the asymptotic radial derivatives, while the factor $f$ keeps the second-derivative combination regular through the horizon. From these numerical plateaus we extract\footnote{After the public release of our numerical results, Villatoro~\cite{villatoro_2026_22011858} independently obtained the leading stress tensor of a massless minimally coupled scalar field using a Kasner-limit calculation. In a common normalization, each of its three independent diagonal stress coefficients differs from our numerical result by less than $0.6\%$, providing an independent check of these leading coefficients.}
\begin{equation}
 \begin{aligned}
 (\lambda_t,\lambda_r,\lambda_\perp)
   &\approx(2.636,-1.193,3.349)\times10^{-3},\\
 q&\approx7.178\times10^{-4}.
 \end{aligned}
 \label{eq:measured}
\end{equation}
The symbols denote asymptotic coefficients; their quoted values and derived numerical predictions are approximate. In the last radial decade, the endpoint-normalized drift of $X,Y,Z,r^3w$ is below $10^{-3}$ (SM Fig.~7), measuring the approach to the asymptotic regime separately from mode-convergence errors.

Inside the horizon $r$ is timelike. In the orthonormal frame $e_{\hat0}=\sqrt{-f}\,\partial_r$, $e_{\hat1}=(-f)^{-1/2}\partial_t$, and normalized angular directions, the numerical plateaus determine the following leading tensor at arbitrary fixed $\xi$ for both computed states $\mathsf{s}\in\{\mathrm U,\mathrm{HH}\}$:
\begin{equation}
 \begin{aligned}
 T^{(\xi,\mathsf{s})}_{\hat a\hat b}
   &=r^{-6}C_{\hat a\hat b}(\xi)+o(r^{-6}),\\
 \begin{pmatrix}C_{\hat0\hat0}\\C_{\hat1\hat1}\\C_{\hat2\hat2}=C_{\hat3\hat3}\end{pmatrix}
   &=\begin{pmatrix}-\lambda_r\\\lambda_t\\\lambda_\perp\end{pmatrix}
     -\xi q\begin{pmatrix}9\\15\\24\end{pmatrix}.
 \end{aligned}
 \label{eq:tensor}
\end{equation}
All leading off-diagonal entries vanish. Equation~\eqref{eq:tensor} collects the full coupling dependence of the numerical result; the values entering it are given in Eq.~\eqref{eq:measured}.

Let $\rho_q$, $p_\parallel$, and $p_\perp$ denote the frame energy density and principal pressures, with coefficients $c_\rho=C_{\hat0\hat0}$, $c_\parallel=C_{\hat1\hat1}$, and $c_\perp=C_{\hat2\hat2}$. For $c_\rho\ne0$, their limiting ratios $\gamma_i(\xi)=\lim_{r\to0}p_i/\rho_q$ give the anisotropic equation of state
\begin{equation}
 \begin{aligned}
 \gamma_\parallel(\xi)
   &=\frac{\lambda_t-15\xi q}{-\lambda_r-9\xi q},\\
 \gamma_\perp(\xi)
   &=\frac{\lambda_\perp-24\xi q}{-\lambda_r-9\xi q}.
 \end{aligned}
 \label{eq:eos}
\end{equation}
Here equation of state means an asymptotic relation between the energy density and principal pressures of the mean stress. At minimal coupling, $\gamma_\parallel\approx2.21$ and $\gamma_\perp\approx2.81$. At each fixed $\xi$, the Unruh and Hartle--Hawking states have the same limiting ratios and density coefficient. At $\xi_\rho=-\lambda_r/(9q)\approx0.185$, the leading density vanishes: the ratios cease to apply, while the coefficient tensor remains well defined.

Figure~\ref{fig:basis}(b) quantifies the convergence of the two states to the same diagonal-stress and scalar plateaus near the singularity, despite their distinct exterior profiles. Figure~\ref{fig:basis}(a) also directly shows that the common leading tensor coexists with different net energy transport: $T^r{}_t\approx-5.920\times10^{-6}/r^2$ in the Unruh state, whereas the Hartle--Hawking flux vanishes. In the interior orthonormal frame defined above, the Unruh flux component is
\begin{equation*}
 T^{(\xi,\mathrm U)}_{\hat0\hat1}
 =\frac{T^r{}_t}{f}
 \approx\frac{2.960\times10^{-6}}{r},\qquad r\to0.
\end{equation*}
It is subleading to the $r^{-6}$ energy density and pressures. The leading quantum source is therefore dominated by vacuum-polarization stress. This numerical agreement between the Unruh and Hartle--Hawking states raises a broader question: does the leading singular stress change when the quantum state is varied?

\textit{State independence for general quantum environments.---}
To answer this question, we compare the Unruh state with an arbitrary Hadamard state \(\mathsf{s}\) defined throughout the Schwarzschild interior \(0<r<2\), without assuming stationarity or spherical symmetry of \(\mathsf{s}\). Let $W_{\mathsf{s}}(x,x')=\langle\Phi(x)\Phi(x')\rangle_{\mathsf{s}}$ and $h=W_{\mathsf{s}}-W_{\mathrm U}$; throughout this comparison, $\Delta$ denotes the state difference and the renormalization prescription is held fixed. The Hadamard singularities and common commutator cancel, so $h$ is smooth, real, symmetric, and satisfies $\Box_xh=\Box_{x'}h=0$~\cite{Radzikowski:1996pa}.

Fix an interior slice $0<r_0<2$ and a compact tangential set $K\subset\mathbb R_t\times S^2$. Write $\eta=\log(r_0/r)$ and $L=1+\eta$. Interior causal propagation has finite extent in $t$, so only compact portions of the four initial kernels $H^{ij}=[\partial_\eta^i\partial_{\eta'}^jh]_{\eta=\eta'=0}$, $i,j=0,1$, affect $K$. A smooth cutoff and an auxiliary period $P$ in the spatial coordinate $t$ give smooth kernels $\widetilde H^{ij}$ on $S^1_P\times S^2$. With $y=(t,\Omega)$, $e_p(y)=P^{-1/2}e^{ik_nt}Y_{\ell m}(\Omega)$, $p=(n,\ell,m)$, and $k_n=2\pi n/P$, their expansion obeys
\begin{equation}
 \begin{aligned}
 \widetilde H^{ij}(y,y')&=\sum_{p,q}c^{ij}_{pq}e_p(y)e_q(y'),\\
 |c^{ij}_{pq}|&\le C_N\kappa_p^{-2N}\kappa_q^{-2N},\qquad N\ge0,\\
 \kappa_p^2&=1+k_n^2+\ell(\ell+1),\\
 \sum_p\kappa_p^{-a}&<\infty\quad(a>3).
 \end{aligned}
 \label{eq:spectral}
\end{equation}
The coefficient bound follows by repeated integration by parts in both arguments with the positive elliptic operator
\begin{equation*}
 \mathcal A=1-\partial_t^2-\Delta_{S^2},\qquad
 \mathcal A e_p=\kappa_p^2e_p,
\end{equation*}
where $\Delta_{S^2}$ is the Laplacian on the unit sphere. It holds for every integer $N\ge0$.

With $b=2-r$, the separated radial equation is $(bU_{,\eta})_{,\eta}+\mathcal V_pU=0$, where $\mathcal V_p=\ell(\ell+1)r+k_n^2r^4/b\ge0$. An energy estimate gives
\begin{equation}
 \begin{aligned}
 \mathcal E_p&=b|U_{,\eta}|^2+\mathcal V_p|U|^2,\\
 \partial_\eta\mathcal E_p
   &=-r|U_{,\eta}|^2+(\partial_\eta\mathcal V_p)|U|^2\le0,\\
 |U_{p,i,\eta}|&\le C\kappa_p,\qquad
 |U_{p,i}|\le C\kappa_p L,\\
 |U_{p,i,\eta\eta}|&\le Cr\kappa_p^3L.
 \end{aligned}
 \label{eq:energy}
\end{equation}
The fundamental solutions $U_{p,0}$ and $U_{p,1}$ have initial data $(U,U_{,\eta})=(1,0)$ and $(0,1)$, respectively. All constants are uniform in $p$ and $r$. The logarithmic field growth agrees with classical wave asymptotics~\cite{Fournodavlos:2018lrk}.

Writing $V_{p,i}=U_{p,i}e_p$, reconstruct the two-point-function difference as
\begin{equation}
 \widetilde h(x,x')=\sum_{i,j=0}^{1}\sum_{p,q}
 c^{ij}_{pq}V_{p,i}(x)V_{q,j}(x').
 \label{eq:reconstruction}
\end{equation}
On any compact region with $r,r'>0$, derivatives through second order introduce only polynomial spectral weights. Equation~\eqref{eq:spectral} therefore makes the series and all these derivative series uniformly convergent, allowing either wave operator to act term by term. The reconstructed function solves both wave equations and has the prescribed four initial kernels. Cauchy uniqueness and finite propagation then identify it with $h$ on the target region.

To control the singular limit, the frame derivatives satisfy $|e_{\hat a}V_{p,i}|\le Cr^{-3/2}\kappa_p^3$. Taking $N=4$ in Eq.~\eqref{eq:spectral} bounds the full differentiated sum by
\begin{equation}
 \begin{aligned}
 &(rr')^{3/2}\sum_{i,j,p,q}|c^{ij}_{pq}|\,
       |e_{\hat a}V_{p,i}|\,|e_{\hat b'}V_{q,j}|\\
 &\qquad\le C\left(\sum_p\kappa_p^{-5}\right)^2<\infty .
 \end{aligned}
 \label{eq:sumbound}
\end{equation}
The two frame derivatives act at $x$ and $x'$, respectively. The summable bound is independent of $r,r'$, so the full frequency and angular sums preserve the radial order of the estimate.

At each fixed $r>0$, define
\begin{equation*}
 Q_{ab}=[\Pi_b{}^{b'}\nabla_a\nabla_{b'}h],\qquad
 Q=g^{ab}Q_{ab}.
\end{equation*}
Here $\Pi_b{}^{b'}(x,x')$ is the parallel transport operator that maps a covector from $x'$ to $x$ along their local connecting geodesic, and brackets denote coincidence $x'\to x$. This notation is also known as the Synge bracket. Equation~\eqref{eq:sumbound} gives $Q_{\hat a\hat b}=O_K(r^{-3})$. The minimally coupled stress operator gives $\Delta T^{(0)}_{ab}=Q_{ab}-\tfrac12g_{ab}Q$. Differentiating the convergent series for $\Delta w=[h]$ and projecting its second covariant derivative into the orthonormal frame yields, with Eq.~\eqref{eq:improvement},
\begin{equation}
 \begin{aligned}
 \Delta w&=O_K(L^2),\\
 e_{\hat a}{}^a e_{\hat b}{}^b\nabla_a\nabla_b\Delta w
    &=O_K(r^{-3}L),\\
 \Delta T^{(0)}_{\hat a\hat b}&=O_K(r^{-3}),\\
 \Delta T^{(\xi)}_{\hat a\hat b}
   &=O_{K,\xi}(r^{-3}L)=o_K(r^{-6}).
 \end{aligned}
 \label{eq:comparison}
\end{equation}
Here $e_{\hat a}{}^a$ are the components of the interior orthonormal frame. These estimates are uniform on $K$ for every fixed finite $\xi$; their constants may depend on the state, $r_0$, $K$, and $\xi$, but not on $r\downarrow0$. Only the common causal propagation region is needed. The precise domain conditions and complete estimates are given in SM Sec.~VI~\cite{Supplement}.

The motion of a black hole relative to its surrounding quantum environment, together with the evolution of the environment itself, can make the quantum state depend on direction and time. Describing such environments therefore requires allowing states beyond stationary ones such as Unruh and Hartle--Hawking, or their finite wave-packet excitations. Combining Eq.~\eqref{eq:comparison} with the numerically determined tensor in Eq.~\eqref{eq:tensor}, we establish a \textbf{rigidity theorem for the leading RSET} of a massless scalar field on the fixed Schwarzschild background: for each given curvature coupling $\xi$, all leading coefficients, including their numerical values and signs, are independent of the choice of Hadamard state on the common interior evolution domain specified above.

Villatoro's Kasner-limit analysis~\cite{villatoro_2026_22011858} shows that a smooth one-particle excitation contributes only subleading stress. Our uniform estimates for the complete mode sums control the full two-point-function difference for general Hadamard states, including correlations between different modes. The result does not require the corresponding state to be stationary, as the Unruh and Hartle--Hawking states are, nor does it restrict state changes to finite wave-packet excitations. The theorem therefore applies to any quantum environment described by the Hadamard class specified above.

The leading mean stress consequently inherits the symmetries of the Schwarzschild background, while the time and directional dependence of the quantum environment enters only at subleading order. For a given field theory and curvature coupling, this expresses the rigidity of the leading quantum source governed by the near-singularity geometry.

\textit{Spacelike singularity and inner-horizon blueshift.---}
The subextremal, asymptotically flat Reissner--Nordstr\"om interior provides a concrete contrast~\cite{ZilbermanLeviOri2020,Sela2018,LanirOriZilberman2019}. Let $F_v^{(\mathsf{s})}$ be the limiting renormalized $T_{vv}$ in state $\mathsf{s}$ at the inner horizon, and let $V_-=-e^{-\kappa_-v}$ be a regular null coordinate, with $\kappa_->0$ the inner-horizon surface gravity. Then
\begin{equation}
 T^{(\mathsf{s})}_{V_-V_-}
  =\frac{F_v^{(\mathsf{s})}}{\kappa_-^2V_-^2}
      +o(V_-^{-2}),\qquad V_-\to0.
 \label{eq:chcontrast}
\end{equation}
The different Unruh and Hartle--Hawking values of $F_v$ show that information about the black hole's quantum environment can survive in the leading divergent term near the Cauchy horizon~\cite{ZilbermanLeviOri2020}. This contrasts sharply with the state-independent rigidity of the leading tensor near the Schwarzschild singularity.

The distinction corresponds to different propagation mechanisms: the Cauchy-horizon blueshift occurs at finite background curvature and involves unbounded advanced time; curvature grows without bound at the Schwarzschild singularity, but the past dependence of a fixed compact tangential region approaching it remains uniformly contained in a compact region on an earlier interior slice, independently of the target radius. The contrast here is specific to asymptotically flat Reissner--Nordstr\"om. In Reissner--Nordstr\"om--de Sitter, Ref.~\cite{Hollands:2019whz} compares Hadamard states on an initial-data domain of dependence extending across the event and cosmological horizons and obtains a state-independent leading Cauchy-horizon coefficient in the parameter range studied there.

\textit{Directional null projections.---}
A further consequence of Eq.~\eqref{eq:tensor} is its coupling-dependent null sign structure. Let $\vartheta$ be the angle between the spatial direction of the null vector and the longitudinal frame vector $e_{\hat1}$. At a point normalize $k^{\hat0}=1$, so that $\mu=(k^{\hat1})^2=\cos^2\vartheta\in[0,1]$. Contracting the same numerical tensor gives
\begin{equation}
 \begin{aligned}
 r^6T_{ab}^{(\xi)}k^ak^b&=\mathcal N(\xi,\mu)+o(1),\\
 \mathcal N(\xi,\mu)&=\mu\mathcal N_\parallel+(1-\mu)\mathcal N_\perp,\\
 \mathcal N_\parallel&=\lambda_t-\lambda_r-24\xi q,\\
 \mathcal N_\perp&=\lambda_\perp-\lambda_r-33\xi q.
 \end{aligned}
 \label{eq:projection}
\end{equation}
These coefficients are the leading density-plus-pressure combinations. Through the Einstein--Hilbert semiclassical equation, positive and negative $T_{ab}k^ak^b$ provide focusing and defocusing contributions to the Raychaudhuri equation, respectively. The numerical zeros of $\mathcal N_\perp$ and $\mathcal N_\parallel$ occur at $\xi_\perp\approx0.192$ and $\xi_\parallel\approx0.222$, respectively. The source is positive in all directions below the first threshold, has mixed directional signs between them, and is negative in all directions above the second.

\begin{figure}[!tbp]
\centering
\includegraphics[width=0.94\columnwidth]{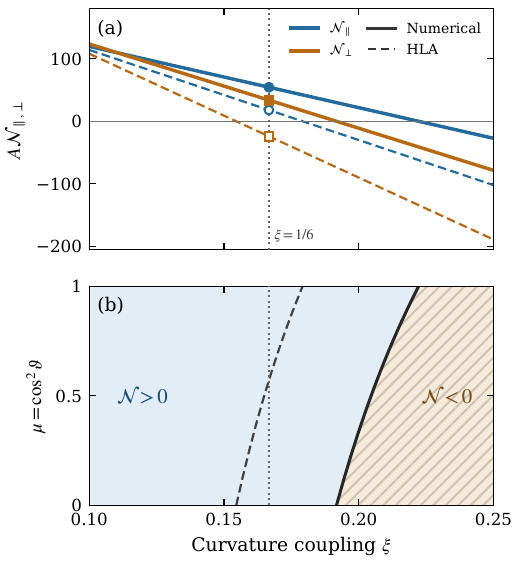}
\caption{Directional null projections derived from the universal tensor. (a) Longitudinal and transverse coefficients multiplied by $A=5760\pi^2$: numerical results (solid) and HLA (dashed); symbols mark conformal coupling. (b) Numerical sign diagram with $\mu=\cos^2\vartheta$, where $\vartheta$ is the angle of the null spatial direction to the longitudinal frame vector; $\mu=1$ is longitudinal and $\mu=0$ transverse. Solid and dashed curves show numerical and HLA zero loci. The dotted line marks $\xi=1/6$. On a zero locus, the next-order sign is not determined by the leading tensor.}
\label{fig:signs}
\end{figure}

Figure~\ref{fig:signs} also compares this consequence with the Hiscock--Larson--Anderson (HLA) approximation~\cite{Hiscock:1997jt}. With $A=5760\pi^2$, the HLA approximation gives $A\mathcal N_\parallel=258-1440\xi$ and $A\mathcal N_\perp=306-1980\xi$. At conformal coupling $\xi=1/6$,
\begin{equation}
 \begin{aligned}
 A(\mathcal N_\parallel,\mathcal N_\perp)_{\rm num}&\approx(54.4,\,33.8),\\
 A(\mathcal N_\parallel,\mathcal N_\perp)_{\rm HLA}&=(18,\,-24).
 \end{aligned}
 \label{eq:conformaltest}
\end{equation}
These results show that the numerically determined leading tensor and the HLA approximation make qualitatively different predictions for the mean null projections and their focusing contributions: at conformal coupling, the numerical tensor provides a focusing contribution in every direction, whereas HLA gives a transverse defocusing contribution. Figure~\ref{fig:signs}(b) further displays the sign diagram in curvature coupling and null direction. A slice at fixed $\xi$ specifies the focusing or defocusing contribution of the mean quantum source along each null direction, providing quantitative input for analyzing directional dependence in semiclassical backreaction. By Eq.~\eqref{eq:comparison}, nonzero leading signs are unchanged throughout the stated Hadamard class; on the zero loci, subleading terms determine the sign.

\textit{Conclusion.---}
By standard point splitting, we directly determine the complete RSET of a massless scalar field at arbitrary curvature coupling in the four-dimensional Schwarzschild interior, extracting its leading coefficient tensor from the common near-singularity plateaus of the Unruh and Hartle--Hawking states. This fixes the energy density, anisotropic pressures, and their coupling-dependent asymptotic ratios. The general Hadamard comparison further proves that, at fixed coupling, changing the quantum state on the common interior evolution domain cannot change these leading coefficients, even when the quantum environment depends on time and direction. This rigidity contrasts sharply with state-sensitive leading Cauchy-horizon fluxes in asymptotically flat Reissner--Nordstr\"om spacetime.

The null-projection diagram characterizes the focusing and defocusing contributions of this universal leading quantum source as curvature coupling and direction vary. At conformal coupling, the numerical result gives positive projections in every direction, in qualitative contrast with the negative transverse projection of HLA. These results provide quantitative input for assessing semiclassical backreaction near a spacelike singularity. Determining the resulting geometry still requires a self-consistent metric response and control of stress fluctuations~\cite{Borgman:2003dm,Hu:2004gf,Perez:2025tvg}. Fluctuation-induced focusing~\cite{Bak:2022oyn,Bak:2023wwo}, stochastic interior evolution~\cite{Wang:2026fql}, and semiclassical singularity constraints~\cite{Bousso2025} address complementary aspects. Subleading stresses and higher correlations can still retain environmental information.

\begin{acknowledgments}
\textit{Acknowledgments.---} S.~J. is supported by the National Natural Science Foundation of China under Grant No.~12275087. J.~J. is supported by the National Natural Science Foundation of China under Grant No.~12205014.
\end{acknowledgments}
\textit{Data availability.---} The numerical data supporting this study are available from the corresponding author upon reasonable request.

\renewcommand{\bibfont}{\fontsize{9}{10.3}\selectfont}

\end{document}